\documentclass[runningheads]{svmult}

\usepackage{makeidx}   %
\usepackage{graphicx}  %
\usepackage{subeqnar}  %
\usepackage{multicol}  %
\usepackage{physprbb}  %
\usepackage{amsfonts} %
\makeindex             %

\usepackage{epsfig}

\begin{document}
\title*{The Two-Time Green's Function and Screened Self--Energy for
  Two-Electron Quasi-Degenerate States}
\toctitle{The Two-Time Green's Function and
\protect\newline Screened Self--Energy for Two-Electron Quasi-Degenerate States}
\titlerunning{Atomic Energy Levels with QED}
\author{\'Eric-Olivier Le~Bigot\inst{1}
\and Paul Indelicato\inst{1}
\and Vladimir M. Shabaev\inst{2}}
\authorrunning{\'Eric-Olivier Le~Bigot et al.}
\institute{
Laboratoire Kastler-Brossel, Case 74,
\'ENS et Universit\'e P. et M. Curie\\
Unit\'e Mixte de Recherche du CNRS n$^\circ$ C8552\\
4, pl.~Jussieu, 75252 Paris CEDEX 05, France
\and
Department of Physics,
St. Petersburg State University\\
Oulianovskaya 1, Petrodvorets, St. Petersburg 198904, Russia}

\maketitle              %

\label{c_lebi}

\index{Exact calculations!in Zalpha@in $Z\alpha$|(}

\begin{abstract}
  Precise predictions of atomic energy levels require the use of QED,
  especially in highly-charged ions, where the inner electrons have
  relativistic velocities. We present an overview of the two-time
  Green's function method; this method allows one to calculate level
  shifts in two-electron highly-charged ions by including in principle
  all QED effects, for any set of states (degenerate, quasi-degenerate
  or isolated). We present an evaluation of the contribution of the
  screened self-energy to a finite-sized effective hamiltonian that
  yields the energy levels through diagonalization.
\end{abstract}

\section{Experiments and Theory}
Experimental measurements of atomic energy levels provide more and
more stringent tests of theoretical models; thus, the experimental
accuracy of many measurements is better than the precision of
theoretical calculations: in \emph{hydrogen}\index{Hydrogen atom}%
~\cite{51debeauvoir97,51huber99}, in \emph{helium}~\cite{51dorrer97,51drake96}%
\index{Helium atom}, and in \emph{lithium}-like uranium%
\index{Lithium-like ion!uranium}~\cite{51schweppe91} and bismuth%
\index{Lithium-like ion}~\cite{51beiersdorfer98}. The current status of
many precision tests of Quantum-Electrodynamics in hydrogen and helium
can be found in this edition.

Furthermore, highly-charged ions possess electrons that move with a
velocity which is close to the speed of light.  \index{Few-electron
  ions}\index{High-Z ions@High-$Z$ ions} The theoretical study of such
systems must therefore take into account \emph{relativity}; moreover,
a \emph{perturbative} treatment of the binding to the \emph{nucleus}
(with coupling constant $Z\alpha$) fails in this
situation~\cite{51yerokhin00}%
\index{Zalpha expansion@$Z\alpha$ expansion}. Perturbative expansions
in $Z\alpha$, however, are useful in different situations (see
\cite{51pachucki98b} for a review, and articles in {\em this
edition}
\cite{51karshenboim2000,51melnikov2000,51andreev2000,51ivanov2000}).

\section{Theoretical Methods for Highly-Charged Ions}

\index{Few-electron ions}\index{High-Z ions@High-$Z$ ions} There are
only a few number of methods that can be used in order to predict
energy levels for \emph{highly-charged ions} within the framework of
Bound-State Quantum Electrodynamics~\cite{51mohr89}\index{Bound state
  QED}: the adiabatic $S$-matrix \index{S-matrix} formalism of
Gell-Mann, Low and Sucher~\cite{51sucher57}, the evolution operator
method~\cite{51vasilev75,51zapryagaev85}, the two-time Green's function
method~\cite{51shabaev94} and an interesting method recently proposed by
Lindgren (based on Relativistic Many-Body Perturbation Theory merged
with QED)~\cite{51lindgren00}. All these methods are based on a study of
the some evolution operator or propagator; the two extreme times of
the propagation can be \emph{both infinite} (Gell-Mann--Low--Sucher),
\emph{one} can be finite and the other infinite (Lindgren), and both
can be finite (Shabaev).

But among these methods, only \emph{two} can in principle be used in
order to apply perturbation theory to \emph{quasi-degenerate} levels
(e.g., the $^3$P$_1$ and $^1$P$_1$ levels in helium-like ions): the
two-time Green's function method and Lindgren's method (which is still
under development). \index{Effective Shr\"odinger equation}Both work
by constructing a finite-sized \emph{effective hamiltonian} whose
eigenvalues give the energy levels~\cite{51shabaev93}.

The two-time Green's function method has the advantage of being
applicable to many atomic physics problems, such as the recombination
of an electron with an ion~\cite{51shabaev94b}, the shape of spectral
lines~\cite{51shabaev91} and the effect of nuclear recoil on atomic
energy levels~\cite{51shabaev98,51shabaev2000b}.

\subsection{Overview of the Two-Time Green's Function Method}

\begin{figure}[h]
\begin{center}

  \begin{center} 
  {\resizebox{0.5\textwidth}{!}
      {\epsfig{file=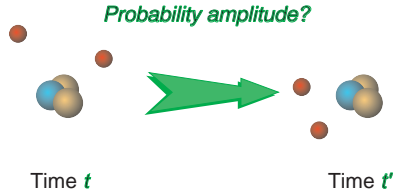}}}
  
  \end{center}
  
\end{center}
\caption[]{The 2-particle Green's function is the amplitude for going
  from one state of two particles to another state
\label{fig:greenDef}
}
\end{figure}

We give in this section a short outline of the two-time Green's
function method. The basic object of this method~\cite{51shabaev90}
represents the probability amplitude for $N$ fermions to go from one
position to the other, as shown in Fig.~\ref{fig:greenDef}.

The corresponding mathematical object is a usual $N$-particle
correlation function between \emph{two} times:
\begin{eqnarray}\label{51eq:fullFProp}
\label{eq:defPropTime}
\lefteqn{{S_{\mbox{\scriptsize{F}}}^N}^{\alpha_1  \ldots  \alpha_N}_{\alpha_1' \ldots
  \alpha_N'}
({\mathbf{x}}_1,\ldots,{\mathbf{x}}_N,t;
  {\mathbf{x}}'_1,\ldots,{\mathbf{x}}'_N, t')}&&\\
&\equiv&
\langle \Omega|\mbox{T}\,
\hat{\Psi}^{\alpha_1}({\mathbf{x}}_1,t)\cdots\hat{\Psi}^{\alpha_N}({\mathbf{x}}_N,t)
\nonumber\\
&&\qquad\,\times\hat{\overline{\Psi}}_{\alpha_N'}
({\mathbf{x}}'_N,t')\cdots\hat{\overline{\Psi}}_{\alpha_1'}({\mathbf{x}}'_1,t')
|\Omega\rangle
\,,
\end{eqnarray}
where~$|\Omega\rangle$ is the vacuum of the \emph{full}
Bound-State QED Hamiltonian~$\hat{H}$, and where the
quantum field~$\hat{\Psi}$ is defined as the usual canonical
electron--positron field evolving under the total hamiltonian in the
\emph{Heisenberg} picture~\cite{51mohr89}.

A remark can be made here about \emph{Lorentz invariance}: the above
correlation function (or propagator) displays only \emph{two times},
which are associated to \emph{many different positions}. A Lorentz
transform of the space--time positions involved therefore yields many
different individual times (one for each position); thus, the object
(\ref{51eq:fullFProp}) must be defined in a \emph{specific} reference
frame. And this reference frame is chosen as nothing more than the
Galilean reference frame associated to the nucleus, which is
physically privileged.

\subsection*{Fundamental Property of the Green's Function}

\begin{figure}[htb]
\begin{center}

  \begin{center} 
  {\resizebox{0.7\textwidth}{!}
      {\epsfig{file=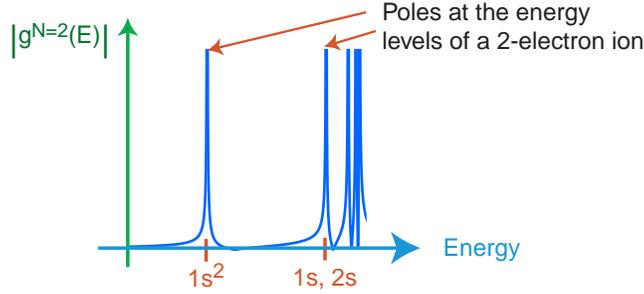}}}
  
  \end{center}
  
\end{center}
\caption[]{The 2-particle Green's function contains information about
  the atomic energy levels of a 2-electron atom or ion
\label{fig:polesGreen}
}
\end{figure}

The $N$-particle Green's function is a function of \emph{energy}
simply defined through a Fourier transform of Eq.\ (\ref{eq:defPropTime}):
\begin{eqnarray}
\lefteqn{
{\mathcal{G}}^{N}({\mathbf{x}}_1,\ldots,{\mathbf{x}}_N; {\mathbf{x}}'_1,\ldots,{\mathbf{x}}'_N; E \in {\mathbb R})
}\nonumber\\
\label{51eq:defGreenInit}
&\equiv&\frac{1}{i}
\int \! d\Delta t\,
e^{iE\Delta t}\,
S_{\mbox{\scriptsize{F}}}^N({\mathbf{x}}_1,\ldots,{\mathbf{x}}_N,\Delta t; {\mathbf{x}}'_1,\ldots,{\mathbf{x}}'_N,t'=0)
\,.
\end{eqnarray}
This function is interesting because it contains the energy levels
predicted by Bound-State QED: one can show~\cite{51shabaev90} that
\begin{eqnarray}
\label{51eq:greenZerothOrder}
\lefteqn{
{\mathcal{G}}^{N}({\mathbf{x}}_1,\ldots,{\mathbf{x}}_N; {\mathbf{x}}'_1,\ldots,{\mathbf{x}}'_N; E \in {\mathbb R})
}\\
&=&
\sum_{\mbox{\scriptsize{\parbox{11em}{\centering
      Eigenstates $|n\rangle$ of $\hat{H}$ with charge~$-N|e|$}}}}
\frac{\langle \Omega|{
\hat{\psi}({\mathbf{x}}_1) \cdots \hat{\psi}({\mathbf{x}}_N)
|n\rangle\langle n|
\hat{\overline{\psi}}({\mathbf{x}}'_N)\cdots \hat{\overline{\psi}}({\mathbf{x}}'_1)
}|\Omega\rangle}
{E-(E_n-i0)}\nonumber\\
\nonumber
&
{}+(-1)^{N^2+1}
&
\sum_{\mbox{\scriptsize{\parbox{11em}{\centering
      Eigenstates $|n\rangle$ of $\hat{H}$ with charge~$+N|e|$}}}}
\frac{
\langle \Omega|{
\hat{\overline{\psi}}({\mathbf{x}}'_N) \cdots \hat{\overline{\psi}}({\mathbf{x}}'_1)
|n\rangle\langle n|
\hat{\psi}({\mathbf{x}}_1)\cdots \hat{\psi}({\mathbf{x}}_N)
}|\Omega\rangle
}
{E-(-E_n+i0)}
\,,
\end{eqnarray}
\noindent
where $|\Omega\rangle$ is the vacuum of the total hamiltonian
$\hat{H}$; $\hat{\psi}$ is the usual second-quantized Dirac field in
the Schr\"odinger representation and $E_n$ is the energy of the
eigenstate $n$ of $\hat{H}$.  The poles in $E$ with a positive real
part are exactly the energies of the states with charge $-N|e|$,
which are physically the atomic eigenstates of an ion with $N$
orbiting electrons (The charge of the nucleus is not counted in the
total charge.), as shown graphically in Fig.~\ref{fig:polesGreen}. 
Such a result is similar to the so-called 
K\"all\'en--Lehmann representation~\cite{51peskin95}.

In order to obtain the energy levels contained in
(\ref{51eq:greenZerothOrder}), we must resort on a perturbative
calculation of the correlation function (\ref{51eq:fullFProp}), which
belongs to standard textbook knowledge~\cite{51itzykson}. The position
of the \emph{poles} of (\ref{51eq:greenZerothOrder}) must then be
mathematically found. It is possible to construct an effective,
finite-size hamiltonian which acts on the atomic state that one is
interested in; the eigenvalues of this hamiltonian then give the
Bound-State QED evaluation of the energy levels~\cite{51shabaev93}. This
hamiltonian is obtained through contour integrations.

\subsection{Second-Order Calculations}

\begin{figure}[h]
\begin{center}

  \begin{center} 
  {\resizebox{0.9\textwidth}{!}
      {\epsfig{file=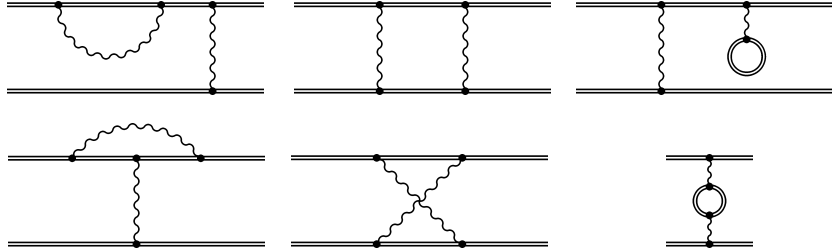}}}
  
  \end{center}
  
\end{center}
\caption[]{
\label{fig:diagsScndOrder}
The contributions of
  order $\alpha^2$ to the electron-electron interaction}
\end{figure}

The current state-of-the-art in non-perturbative calculations (in
$Z\alpha$) of atomic energy levels within Bound-State QED consists in
the theoretical evaluation of the contribution of diagrams with
\emph{two photons} ({i.e.} of order $\alpha^2$, since the
electron--photon coupling constant is $e$).
\index{Radiative corrections}
For instance, for ions with  two electrons, the screening of
one electron by the other is described by the six
diagrams of Fig.~\ref{fig:diagsScndOrder}.

However, most of the calculations of contributions of order $\alpha^2$
were, until very recently, restricted to the very specific case of
the ground-state (see \cite{51yerokhin99} for references). The extension
to the calculation of the energy levels of quasi-degenerate states
represents one of the current trends of the research in the domain of
non-perturbative (in $Z\alpha$) calculations with QED.

We have calculated the contribution of the screened self-energy (first
and fourth
diagrams of Fig.~\ref{fig:diagsScndOrder}) to some isolated levels in
\cite{51yerokhin99,51indelicato98a,51indelicato91,51yerokhin99b}. When
energy levels are quasi-degenerate (e.g., the $^3$P$_1$ and $^1$P$_1$
levels in helium-like ions), the two-time Green's function method
allows one to evaluate the matrix elements of the effective
hamiltonian between different states;
for the first diagram of Fig.~\ref{fig:diagsScndOrder}, we obtain
the following contribution to this hamiltonian (The two
electrons on the left are denoted by $n_1$ and $n_2$, and the two on
the right by $n'_1$ and $n'_2$, and other notations follow.):
\begin{eqnarray}
\nonumber
&&
\sum_{P,P'} \ensuremath{(-1)^{PP'}}
\Bigg\{
-
\Big(
\sum_{k \ne n'_{P'(1)}}
\mbox{$\langle n_{P(1)} n_{P(2)}|$}S_{k}^\mathrm{r}(\varepsilon_{n_{P(1)}},\varepsilon_{n'_{P'(1)}})\mbox{$|n'_{P'(1)} n'_{P'(2)}\rangle$}
\\
\label{eq:effHamSSEIrr}
&&
\qquad
+
\sum_{k \ne n_{P(1)}}
\mbox{$\langle n_{P(1)} n_{P(2)}|$}S_{k}^\mathrm{l}(\varepsilon_{n_{P(1)}},\varepsilon_{n'_{P'(1)}})\mbox{$|n'_{P'(1)} n'_{P'(2)}\rangle$}
\Big) 
\\
\nonumber
&&
\quad+
\ensuremath{\frac{1}{2}}
\bigg[
\left.\partial_{p}\right|_{\varepsilon_{n_{P(1)}}}
\Big(
\mbox{$\langle n_{P(1)}|$}\Sigma(p)\mbox{$|n_{P(1)}\rangle$}
\\
\nonumber
&&
\quad\qquad\times
\mbox{$\langle n_{P(1)} n_{P(2)}|$}I(p-\varepsilon_{n'_{P'(1)}})\mbox{$|n'_{P'(1)}
  n'_{P'(2)}\rangle$}
\Big)
\\
\nonumber
&&
\qquad+
\left.\partial_{p'}\right|_{\varepsilon_{n'_{P'(1)}}}
\Big(
\mbox{$\langle n_{P(1)} n_{P(2)}|$}I(\varepsilon_{n_{P(1)}} - p')\mbox{$|n'_{P'(1)}
  n'_{P'(2)}\rangle$}
\\
\nonumber
&&
\quad\qquad\times
\mbox{$\langle n'_{P'(1)}|$}\Sigma(p')\mbox{$|n'_{P'(1)}\rangle$}
\Big)
\bigg]
\Bigg\}
\\
\nonumber
&&
+
\mathcal{O}[\alpha^2 (E_{n'}^{(0)}-E_{n}^{(0)})]
\,,
\end{eqnarray}
where we made use of standard notations~\cite{51yerokhin99}:
$\varepsilon_{k}$ is the energy of the Dirac state $k$, $(-1)^{PP'}$
is the signature of the permutation $P\circ P'$ of the indices
$\{1,2\}$, $\Sigma$ represents the self-energy, and
$I$ represents the photon-exchange:
\begin{eqnarray}
\label{eq:defI}
\mbox{$\langle ab|$}I(\omega)\mbox{$|cd\rangle$}
 &\equiv&
e^2
\int \! d^{3}\vec{x}_2\,
[\psi_a^{\dagger}(\vec{x}_1)\alpha^{\mu}\psi_c(\vec{x}_1)]
\\
\nonumber
&&\times
[\psi_b^{\dagger}(\vec{x}_2)\alpha^{\nu}\psi_d(\vec{x}_2)]
D_{\mu\nu}(\omega; \vec{x}_1-\vec{x}_2)
\\
\label{eq:defSigma}
\mbox{$\langle a|$}\Sigma(p)\mbox{$|b\rangle$}
&\equiv&
\frac{1}{2\pi i} \int \! d\omega \,
\sum_{k}\frac{\mbox{$\langle ak|$}I(\omega)\mbox{$|kb\rangle$}}
{\varepsilon_k(1-i0) - (p-\omega)},
\end{eqnarray}
where $a,b,c$ and $d$ label Dirac states, and $e$ is the charge of the
electron; $\alpha^{\mu}$ are the Dirac matrices, and 
$\psi$ denotes a Dirac spinor; the photon propagator $D$
is given in the Feynman gauge by:
\begin{equation}
D_{\nu\nu'}(\omega; \vec{r})
 \equiv 
g_{\nu\nu'} \frac{\exp\left(i|\vec{r}|\sqrt{\omega^2-\mu^2+i
      0}
  \right)}
{4\pi|\vec{r}|},
\end{equation}
where $\mu$ is a small photon mass that eventually tends to zero, and
where the square root branch is chosen such as to yield a decreasing
exponential for large real-valued energies $\omega$.  In
Eq.\ (\ref{eq:effHamSSEIrr}), $\partial_x\left|_{x_0}\right.$ is the partial
derivative with respect to $x$ at the point $x_0$, and the
\emph{skeletons} of the screened self-energy diagrams with a
self-energy on the left and on the right are defined as:
\begin{eqnarray*}
\lefteqn{
\mbox{$\langle n_{P(1)} n_{P(2)}|$}S_{k}^\mathrm{r}(p,p')\mbox{$|n'_{P(1)} n'_{P(2)}\rangle$}
\equiv
}
&
&
\\
&&
\qquad
\mbox{$\langle n_{P(1)} n_{P(2)}|$}I(p-p')\mbox{$|k n'_{P'(2)}\rangle$}
\frac{1}{\varepsilon_{k}(1-i0)-  p'}
\mbox{$\langle k|$}\Sigma(p')\mbox{$|n'_{P'(1)}\rangle$}
\,,
\\
\lefteqn{
\mbox{$\langle n_{P(1)} n_{P(2)}|$}S_{k}^\mathrm{r}(p,p')\mbox{$|n'_{P(1)} n'_{P(2)}\rangle$}
\equiv
}
\\
&&
\qquad
\mbox{$\langle n_{P(1)}|$}\Sigma(p)\mbox{$|k\rangle$}
\frac{1}{\varepsilon_{k}(1-i0) - p}
\mbox{$\langle k n_{P(2)}|$}I(p-p')\mbox{$|n'_{P(1)} n'_{P(2)}\rangle$}\,.
\end{eqnarray*}
The terms of order $\alpha^2
(E_{n'}^{(0)}-E_{n}^{(0)})$ are not included in the
above expression because they do not contribute to the level shift of
order $\alpha^2$ in which we are interested. (They contribute to
higher orders, as can be seen in the particular case of two
levels~\cite[p.~27]{51shabaev00}.)

This expression is only formal and must be
renormalized~\cite{51yerokhin99}; angular integrations can then be done
and numerical computations can be performed in order to yield the
Bound-State QED evaluation of the energy shifts.

For the contribution of the first diagram of Fig.~\ref{fig:diagsScndOrder} to
have any physical meaning, it is necessary to calculate it together
with the vertex correction (fourth diagram of Fig.~\ref{fig:diagsScndOrder}).
We have obtained the following contribution to the effective
hamiltonian  for the vertex correction:
\begin{eqnarray*}
\lefteqn{
\sum_{P,P'} \ensuremath{(-1)^{PP'}} \sum_{i_1,i_2} \mbox{$\langle i_1 n_{P(2)}|$}I(\varepsilon_{n_{P(1)}} - \varepsilon_{n'_{P'(1)}})\mbox{$|i_2
  n'_{P'(2)}\rangle$}
}&&
\\
&&
\quad\times
\frac{i}{2\pi} \int \! d\omega \, \frac{\mbox{$\langle n_{P(1)}
    i_2|$}I(\omega)\mbox{$|i_1 n'_{P'(1)}\rangle$}}{
[\varepsilon_{i_1}(1-i0)-(\varepsilon_{n_{P(1)}} - \omega)]
[\varepsilon_{i_2}(1-i0)-(\varepsilon_{n'_{P'(1)}} - \omega)]
}
\\
&&
+\mathcal{O}[\alpha^2 (E_{n'}^{(0)}-E_{n}^{(0)})]
\end{eqnarray*}
where $(n_1, n_2)$ and $(n'_1, n'_2)$ still represent the electrons of
the two states that define the hamiltonian matrix \emph{element} given
here, and where the sum over $i_1$ and $i_2$ is over all Dirac states.

\section{Conclusion and Outlook}

We have presented a quick overview of the current status of
theoretical predictions of energy levels in highly-charged ions with
Bound-State Quantum Electrodynamics. We have given a short description
of the two-time Green's function method, which permits the calculation
of
an effective hamiltonian that can in principle include all QED effects
in energy shifts. We have also presented the specific contribution of
the screened self-energy in the general case (isolated levels,
quasi-degenerate or degenerate levels); the expression obtained can
serve as a basis for numerical calculations of the corresponding
effective hamiltonian.

\index{Exact calculations!in Zalpha@in $Z\alpha$|)}

\providecommand{\bysame}{\leavevmode\hbox to3em{\hrulefill}\thinspace}

\label{c_lebi_}

\end{document}